\documentclass[lettersize,journal]{IEEEtran}
\usepackage{amsmath,amsfonts}
\usepackage{array}
\usepackage[caption=false,font=normalsize,labelfont=sf,textfont=sf]{subfig}
\usepackage{textcomp}
\usepackage{stfloats}
\usepackage{url}
\usepackage{verbatim}
\usepackage{graphicx}
\usepackage{citesort}
\usepackage{balance}
\usepackage{scalerel}
\usepackage{float}
\usepackage{bm}
\usepackage{xcolor}
\usepackage{algorithm}
\usepackage{algorithmicx}
\usepackage{algpseudocode}
\usepackage{amsmath}
\begin{document}

\title{Channel Estimation for Reconfigurable Intelligent Surface-Aided Multiuser Communication Systems Exploiting Statistical CSI of Correlated RIS-User Channels}

        \author{Haochen~Li,~\IEEEmembership{Graduate Student Member,~IEEE,}
        Pan~Zhiwen,~\IEEEmembership{Member,~IEEE,}
			 Wang~Bin,
		   Liu~Nan,~\IEEEmembership{Member,~IEEE,}
        and~You~Xiaohu,~\IEEEmembership{Fellow,~IEEE}
\thanks{Copyright (c) 20xx IEEE. Personal use of this material is permitted. However, permission to use this material for any other purposes must be obtained from the IEEE by sending a request to pubs-permissions@ieee.org. This work is partially supported by the National Key Research and Development Project under Grant 2020YFB1806805 and Science and Technology on Communication Networks Laboratory. (Corresponding author: Pan Zhiwen and Wang Bin.)}
\thanks{Haochen Li, Pan Zhiwen and You Xiaohu are with the National Mobile Communications Research Laboratory, Southeast University, Nanjing 210096, China, and also with the Purple Mountain Laboratories, Nanjing 211100, China (e-mail: lihaochen@seu.edu.cn; pzw@seu.edu.cn; xhyou@seu.edu.cn).}
\thanks{Liu Nan are with the National Mobile Communications Research Laboratory, Southeast University, Nanjing 210096, China (e-mail: nanliu@seu.edu.cn).}
\thanks{Wang Bin is with Science and Technology on Communication Networks Laboratory, Shijiazhuang, Hebei, China (e-mail: 869511890@qq.com).}}

\markboth{Journal of \LaTeX\ Class Files,~Vol.~14, No.~8, August~2021}%
{Shell \MakeLowercase{\textit{et al.}}: A Sample Article Using IEEEtran.cls for IEEE Journals}


\maketitle

\begin{abstract}
Reconfigurable intelligent surface (RIS) is a promising candidate technology for the upcoming Sixth Generation (6G) communication system for its ability to manipulate the wireless communication environment by controlling the coefficients of reflection elements (REs). However, since the RIS usually consists of a large number of passive REs, the pilot overhead for channel estimation in the RIS-aided system is prohibitively high. In this paper, the channel estimation problem for a RIS-aided multi-user multiple-input-single-output (MISO) communication system with clustered users is investigated. First, to describe the correlated feature for RIS-user channels, a beam domain channel model is developed for RIS-user channels. Then, a pilot reuse strategy is put forward to reduce the pilot overhead and decompose the channel estimation problem into several subproblems. Finally, by leveraging the correlated nature of RIS-user channels, an eigenspace projection (EP) algorithm is proposed to solve each subproblem respectively. Simulation results show that the proposed EP channel estimation scheme can achieve accurate channel estimation with lower pilot overhead than existing schemes. 
\end{abstract}


\section{Introduction}
\IEEEPARstart{T}{he} upcoming Sixth Generation (6G) wireless communication networks are expected to impose more stringent requirements on spectrum and energy efficiency. To cope with these increasing demands, the reconfigurable intelligent surface (RIS) has received more and more attention. Via shaping the reflection of impinging signals with adjustable reflection elements (REs), RIS is capable of actively controlling the radio propagation environment and providing unprecedented spectral and energy efficiency increment~\cite{renzo2019smart}. \\
\indent Obtaining channel state information (CSI) is an important prerequisite to take full advantage of RIS-aided systems~\cite{wu2019towards}. Yet, training overhead for channel estimation scales linearly both with the number of RIS REs and that of users. With RIS comprised of a large number of passive REs, the pilot overhead for channel estimation becomes prohibitively high. How to effectively reduce the pilot length is one of the bottlenecks for further implementation of RIS-aided communication systems.
\subsection{Prior Works}
Recently, there are some literature investigating the channel estimation for RIS aided communication systems~\cite{jensen2020optimal,10053657,wang2020compressed,wei2021channel,wang2020channel,zheng2019intelligent,zheng2020intelligent,elbir2020deep}. The authors in~\cite{jensen2020optimal} propose a least square (LS) channel estimation scheme and achieve a minimum variance unbiased estimate of the cascaded channel. {The spatial channel sparsity is leveraged to reduce the pilot overhead using the compressive sensing (CS) technique for RIS-aided systems operating at high-frequency bands in~\cite{wei2021channel,10053657,wang2020compressed}. The authors in~\cite{wang2020channel} propose to estimate the cascaded channel of a reference user at first, and then estimate cascaded channel of other users with reduced pilot overhead. The RIS-aided orthogonal frequency division multiplexing (OFDM) system is considered in~\cite{zheng2019intelligent,zheng2020intelligent}. The authors in~\cite{zheng2019intelligent} propose to use a reflection pattern at the RIS to aid the channel estimation. The authors in~\cite{zheng2020intelligent} exploit the redundancy of OFDM sub-carriers to support more users for concurrent channel estimation and thus improve the training efficiency.} Authors in~\cite{elbir2020deep} design a twin convolutional neural network (CNN) architecture and feed it with the received pilot signals to estimate cascaded channels. However, the quasi-static nature of channels between the base station (BS) and RIS is not utilized in~\cite{jensen2020optimal,10053657,wang2020compressed,wei2021channel,wang2020channel,zheng2019intelligent,zheng2020intelligent,elbir2020deep}.\\ 
\indent Specifically, the BS and RIS are usually fixed and deployed above surroundings, which makes RIS-BS channels vary slowly and easier to be estimated. Thus, it is reasonable to put the main effort into estimating the channels between RIS and users~\cite{hu2021two,he2019cascaded}. To exploit the quasi-static nature of the RIS-BS channels,  authors in~\cite{hu2021two} propose a two-timescale channel estimation framework and use a dual-link pilot transmission scheme to estimate the BS-RIS channel. The authors in~\cite{he2019cascaded} propose a matrix calibration-based channel estimation method to reduce the signaling overhead. \\
\indent Note that the existing works have not considered the spatial correlation between users' antennas and RIS REs. However, due to the insufficient angular spread of the scattering environment and closely spaced antennas/reflecting elements, the channels between RIS and users should be modeled as spatially correlated Rician fading channels~\cite{tsilipakos2020toward}. For spatially correlated Rayleigh fading RIS-user channels, the physical propagation environment makes a few directions more probable to carry strong signals~\cite{bjornson2020rayleigh}, so it is simpler to conduct channel estimation on the subspace spanned by these directions.
\subsection{Contributions}
Against this background, the channel estimation problem for a RIS-aided multi-user multiple-input-single-output (MISO) communication system with clustered users is investigated in this paper. First, to describe the correlated feature for RIS-user channels, the beam domain channel model is developed for RIS-user channels. Then, the pilot reuse strategy is put forward to reduce the pilot overhead and help decompose the channel estimation problem into several subproblems. Finally, by leveraging the correlated nature of RIS-user channels, an eigenspace projection (EP) algorithm is proposed with reduced pilot overhead to solve each subproblem respectively. Our major contributions are summarized as follows:
\begin{enumerate}
\item{In realistic outdoor wireless propagation environments, the RIS is located at an elevated position, and most of the channel power lies in a limited number of spatial directions compared with the whole RIS-user channel dimension. To describe this feature for correlated RIS-user channels, we develop the beam domain channel model for RIS-user channels inspired by~\cite{sun2015beam}.}
\item{For users in different clusters, it is reasonable to assume that angle of arrival (AoA) intervals of RIS-user channels  are non-overlapping. In this circumstance, a pilot reuse scheme is proposed to reduce the pilot overhead.}
\item{Exploiting statistical CSI of correlated channels between RIS and users as a prior, an eigenspace projection (EP) algorithm is proposed to estimate RIS-user channels with reduced pilot overhead.}
\end{enumerate}
\subsection{Organization }
The remainder of this paper is organized as follows: Section II presents the system model of the RIS-aided multi-user MISO system. Section III presents present the channel models for the RIS-BS channel and RIS-user channels. Section IV presents propose the pilot reuse scheme. In Section V, the EP algorithm to estimate RIS-user channels is proposed. Section VI analysis the pilot overhead of our proposed scheme and compares it with different channel estimation schemes. Section VII shows simulation results and analysis. Finally, conclusions are drawn in Section VIII. A list of symbols that are used throughout this article is provided in Table~\ref{Symbols}.
\begin{table*}[h]
\begin{scriptsize}
\caption{{Symbols Used Frequently Throughout This Article and Their Definitions}}\label{Symbols}
\centering
\begin{tabular}{|c||c||c||c|}
\hline
Symbol & Definition & Symbol & Definition\\
\hline
$M$ & Number of BS antennas & $a_{t,n}$, $\theta_{t,n}$ & Reconfigurable amplitude and phase
of the $n$th RE at time t\\
\hline
$N$ & Number of REs & $\mathcal{S}$ & Eigenspace of the sum channel\\
\hline
$N_\mathrm{H}, N_\mathrm{V}$ & Number of REs per row and per column, respectively & $\mathcal{S}_c$ & Eigenspace where
channel realizations of $\mathbf{h}_{i}^{c}$ resides\\
\hline
$C$ & Number of user clusters & $\left\{\mathbf{d}_{1},\mathbf{d}_{2},\cdots,\mathbf{d}_{E}\right\}$ & Orthonormal bases for eigenspace $\mathcal{S}$
\\
\hline
$K$ & Number of users & $E$ & Dimension of eigenspace $\mathcal{S}$ \\
\hline
$I_c$ & Number of users in cluster $c$ & $E_c$ & Dimension of eigenspace $\mathcal{S}_c$\\
\hline
$T$ & Number of subframes in a uplink pilot transmission frame & $\tau$ & Number of time slots in a uplink pilot transmission subframe\\
\hline
\end{tabular}
\end{scriptsize}
\end{table*}
\subsection{Notations}
Throughout the paper, lowercase letters, lowercase bold letters and capital bold letters (e.g., $a$, $\mathbf{a}$ and $\mathbf{A}$) denote scalars, vectors and matrices, respectively. $\mathbb{C}^{M \times K}$ denotes the $M \times K$ dimensional complex vector space. $(\cdot)^\mathrm{T}$, $(\cdot)^\mathrm{*}$ and  $(\cdot)^\mathrm{H}$ represent the operations of transpose, conjugate and conjugate transpose, respectively. The symbol $ \odot $ denotes the Hadamard product. diag$\left( \mathbf{a} \right)$ denotes the diagonal matrix with $\mathbf{a}$ along its main diagonal. $\mathbf{a}_i$ denotes the $i$th element of $\mathbf{a}$. $\lvert \cdot \rvert$, $\lVert \cdot \rVert_2$ and $\lVert \cdot \rVert_\mathrm{F}$ represent the modulus, the Euclidean norm and the Frobenius norm, respectively. $\text{span}\left(\cdot\right)$ denotes the span of vectors. The distribution of a circularly symmetric complex Gaussian (CSCG) random vector with mean vector $\mathbf{a}$ and covariance matrix $\mathbf{A}$ is denoted by $\mathcal{C N}\left(\mathbf{a}, \mathbf{A}\right)$.  $\mathbb{E}\left\{\cdot\right\}$ represents the statistical expectation operator. $\mathbf{I}$ denotes an identity matrix with appropriate dimensions. Calligraphic letters represent linear space, e.g., $\mathcal{S}$. 
\section{System Model}
\begin{figure}[!htbp]
\centering
\includegraphics[width=3.5in]{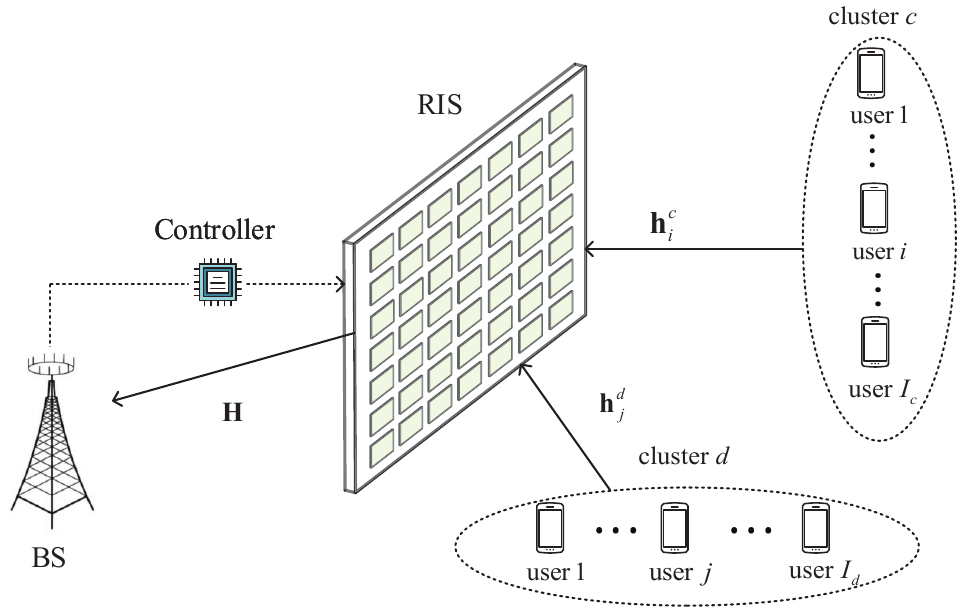}
\caption{An example of RIS-aided wireless communication system.}\label{system_model}
\end{figure}

As shown in Fig.~\ref{system_model}, a narrow-band wireless system, where a RIS is deployed to assist the data transmission between the BS and $K$ users, is investigated. Suppose the RIS is a uniform planer array (UPA) with $N$ REs. {Like works~\cite{9066923,mu2021simultaneously,wang2021joint}, we assume that the direct communication link between the BS and users are blocked by obstacles. This scenario is a classic application for using the RIS to improve the coverage for users in the blind spots~\cite{wu2019towards}.} The BS is equipped with a $M$-antenna uniform linear array (ULA). All users adopt ULA with $M_{\mathrm{u}}$ antennas and locate in $C$ user clusters. The number of users in cluster $c, c\in\left\{1,2,\dots,C\right\}$, is $I_c$, which satisfies $\sum_{c=1}^{C}{{I}_c}=K$.\footnote{User clustering methods have been extensively studied for multi-user communications, such as K-means~\cite{zuo2020resource}, many-to-one matching~\cite{cui2018unsupervised}, correlation of channels~\cite{dai2018hybrid}. These methods can be applied for our considered system.}

Denote
\begin{equation}
\mathbf{\Phi}_{t}=\mathrm{diag}\left(a_{t,1}\exp\left(j\theta_{t,1}\right)  \cdots a_{t,N}\exp\left(j\theta_{t,N}\right)\right)
\end{equation}
as the RIS reflection coefficient matrix at time $t$. In particular, $a_{t,n}$ and $\theta_{t,n}$ represent the reconfigurable amplitude and phase of the $n$th RE in the RIS at time $t$, which can continuously vary within the intervals $\left[0,1\right]$ and $\left[0,2\pi\right)$, respectively.
Then, the received signal of the BS at time $t$ is expressed as
\begin{equation}
\mathbf{y}_t=\sum_{c=1}^{C}\sum_{i=1}^{{I}_c} \mathbf{G} \mathbf{\Phi}_{t} \mathbf{H}_{i}^{c} \mathbf{s}_{ti}^{c} +\mathbf{n}_{t},
\end{equation}
where $\mathbf{G} \in \mathbb{C}^{M \times N}$ denotes the channel from the RIS to the BS, $\mathbf{H}_{i}^{c}\in \mathbb{C}^{N \times M_{\mathrm{u}}}$ represents the channel from the $i$th user in cluster $c$ to the RIS. $\mathbf{s}_{ti}^{c}\in \mathbb{C}^{M_{\mathrm{u}} \times 1}$ represents the transmitted symbol from the $i$th user in cluster $c$ at time $t$. $\mathbf{n}_{t} \sim \mathcal{C N}\left(0, \sigma^{2}\mathbf{I}\right)$ represents the additive white Gaussian noise (AWGN) at time $t$, with $\sigma^{2}$ denoting the noise power.
\section{Channel Model}

\begin{figure}[!htbp]
\centering
\includegraphics[width=3in]{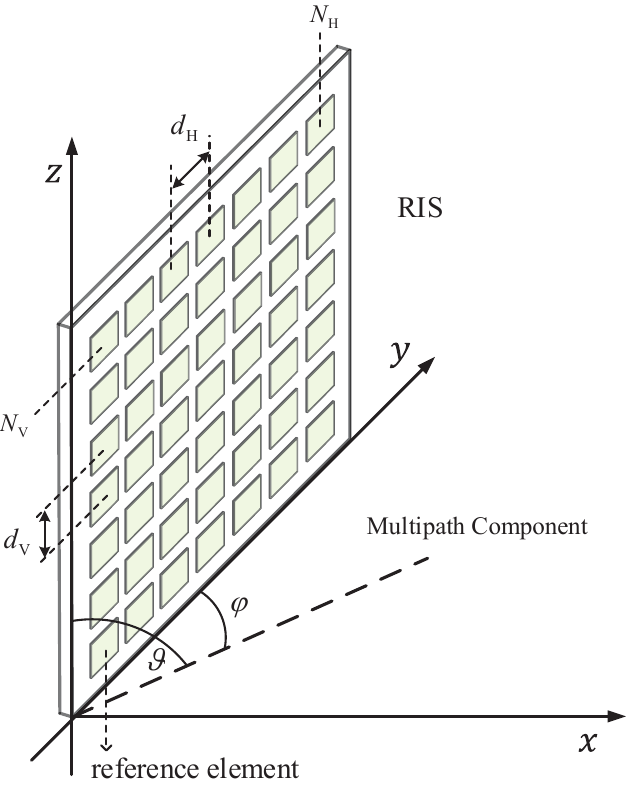}
\caption{The 3D geometry of a rectangular RIS.}\label{RIS}
\end{figure}

The RIS is a rectangular surface consisting of $N_{\mathrm{H}}$ elements per row and $N_{\mathrm{V}}$ elements per column. {The setup is illustrated in Fig.~\ref{RIS} in a three-dimensional(3D) space with $\varphi \in \left[0,\pi\right)$ and $\vartheta \in \left[0,\pi\right)$ being the angles of path to the y-axis and z-axis, respectively.} $d_{\mathrm{H}} = d_{\mathrm{V}} = \lambda/2$ denote the horizontal distance and the vertical distance between neighbouring RIS REs, where $\lambda$ is the wave length. 
The elements are indexed row-by-row by $n \in \left[1,2,\cdots,N\right]$, thus the location of the $n$th element with respect to the origin in Fig.~\ref{RIS} is 
\begin{equation}
\mathbf{u}_{n}=\left[0, i_{\mathrm{H}}\left(n\right)d_{\mathrm{H}}, i_{\mathrm{V}}\left(n\right)d_{\mathrm{V}}\right]^{\mathrm{T}},n\in[1,2,\cdots,N_r]
\end{equation}
where $i_{\mathrm{H}}\left(n\right)$$= \mathrm{mod}\left(n-1,N_{\mathrm{H}}\right)$ and $i_{\mathrm{V}}\left(n\right)$$= \lfloor \left(n-1\right)/N_{\mathrm{H}}\rfloor$ are the horizontal and vertical indices of element n, respectively, on the two-dimensional grid. 

Define $\mathbf{e}_X\left(\cdot\right)$ as the normalized response vector of a ULA with $X$ antennas spaced with $\lambda/2$, $\forall X \in \mathbb{N}^{+}$. The RIS response vector corresponding to $\varphi$ and $\vartheta$ is 
\begin{equation}
\begin{aligned}
\mathbf{e}_\mathrm{RIS}\left(\varphi,\vartheta\right)&=\dfrac{1}{\sqrt{N}}\left[e^{j\mathbf{k}\left(\varphi,\vartheta\right)^{\mathrm{T}}\mathbf{u}_{1}}, \cdots, e^{j\mathbf{k}\left(\varphi,\vartheta\right)^{\mathrm{T}}\mathbf{u}_{N}}\right]^{\mathrm{T}}\\
&=\mathbf{e}_{N_V}\left(\vartheta \right)\otimes\mathbf{e}_{N_H}\left( \varphi\right).
\end{aligned}
\end{equation}
where $\varphi$ and $\vartheta$ represent azimuth AoA and elevation AoA associated with the RIS. $\mathbf{k}\left(\varphi,\vartheta\right) \in \mathbb{R}^{3\times1} $ is the wave vector,
\begin{equation}
\mathbf{k}\left(\varphi,\vartheta\right)=\dfrac{2\pi}{\lambda}\left[0, \cos\left(\varphi\right), \cos\left(\vartheta\right)\right]^{\mathrm{T}}.
\end{equation}
The user response vector corresponding to $\gamma$ is $\mathbf{e}_{M_\mathrm{u}}\left(\gamma\right)$ where $\gamma$ represents angle of departure (AoD) associated with the user.
\subsection{Channel Model for RIS-user Channels and its Spatial Characteristics}

For the RIS and each user, when the sampling of $\gamma_{m}$, $\varphi_{n_\mathrm{H}}$ and $\vartheta_{n_\mathrm{V}}$ satisfies 
\begin{equation}
\varphi_{n_\mathrm{H}}=\arccos(1-2n_\mathrm{H}/N_\mathrm{H}),n_\mathrm{H}\in \left[0,1,\cdots,N_\mathrm{H}\right],
\end{equation}
\begin{equation} 
\vartheta_{n_\mathrm{V}}=\arccos(1-2n_\mathrm{V}/N_\mathrm{V}),n_\mathrm{V}\in \left[0,1,\cdots,N_\mathrm{V}\right],
\end{equation}
and
\begin{equation}
\gamma_m=\arccos(1-2m/ M_\mathrm{u}),m\in \left[0,1,\cdots,M_\mathrm{u}\right],
\end{equation}
the response vectors corresponding to different sampling angles are orthogonal, i.e.,
\begin{equation}
\mathbf{e}_\mathrm{RIS}\left(\varphi_{i_\mathrm{H}\left(n\right)},\vartheta_{i_\mathrm{V}\left(n\right)}\right)^\mathrm{H}\mathbf{e}_\mathrm{RIS}\left(\varphi_{i_\mathrm{H}\left(n'\right)},\vartheta_{i_\mathrm{V}\left(n'\right)}\right)=\delta\left(n-n'\right),
\end{equation} 
\begin{equation}
\mathbf{e}_{M_\mathrm{u}}\left(\gamma_{m}\right)^\mathrm{H}\mathbf{e}_{M_\mathrm{u}}\left(\gamma_{m'}\right)=\delta\left(m-m'\right).
\end{equation}
\indent Suppose that there are $P_i$ physical paths from the RIS to the $i$th user in cluster $c$, and the $p$th path  has an attenuation of $a_{i,p}^{c}$, a distance of $d_{i,p}^{c}$, an angle of $\gamma_{i,p}^{c}$ with the transmit antenna array, and angles of $\varphi_{i,p}^{c}$ and $\vartheta_{i,p}^{c}$  with the receive RE array. Then, the $N\times M_\mathrm{u}$ physical channel matrix associated with the $i$th user in cluster $c$ is given by
\begin{equation}\label{physical_channel_matrix}
\mathbf{H}_{i}^{c}=\sum\limits_{p=1}^{P_i}a_{i,p}^{c}e^{-j2\pi d_{i,p}^{c}/\lambda}\mathbf{e}_\mathrm{RIS}\left(\varphi_{i,p}^{c},\vartheta_{i,p}^{c}\right)\mathbf{e}_{M_\mathrm{u}}^{\mathrm{H}}\left(\gamma_{i,p}^{c}\right).
\end{equation}
\indent It is assumed that users are quite far away from the RIS such that the phases $2\pi d_{i,p}^{c}/\lambda$ are uniformly distributed over $\left[0,2\pi\right)$ and mutually independent due to uncorrelated scattering. These assumptions lead to
\begin{equation}
\mathbb{E}\left\{a_{i,p}^{c}e^{-j2\pi d_{i,p}^{c}/\lambda}\right\}=0,
\end{equation}
\begin{equation}
\begin{aligned}
\mathbb{E}&\left\{a_{i,p}^{c}e^{-j2\pi d_{i,p}^{c}/\lambda}\left(a_{i',p'}^{c}e^{-j2\pi d_{i',p'}^{c}/\lambda}\right)^{*}\right\}
\\&=\beta_{i,p}^{c}\delta\left(i-i',p-p'\right)
\end{aligned}
\end{equation}
where $\beta_{i,p}^{c} = \mathbb{E}\left\{\lvert a_{i,p}^{c} \rvert^{2}\right\}$ is the channel gain of path $p$ to the $i$th user in cluster $c$. 

The channel matrix in \eqref{physical_channel_matrix} can be rewritten as
\begin{equation}
\begin{aligned}
\mathbf{H}_{i}^{c} &= \sum\limits_{m=1}^{M_\mathrm{u}}\sum\limits_{n=1}^{N}\left[\tilde{\mathbf{H}}_{i}^{c}\right]_{nm}\mathbf{e}_{\mathrm{RIS}}\left(\varphi_{i_\mathrm{H}\left(n\right)},\vartheta_{i_\mathrm{V}\left(n\right)}\right)\mathbf{e}_{M_\mathrm{u}}^{\mathrm{H}}\left(\gamma_{m}\right)
\\&= \mathbf{V}\tilde{\mathbf{H}}_{i}^{c}\mathbf{U}^\mathrm{H},
\end{aligned}
\end{equation}
where $\mathbf{V}=\mathbf{V}_\mathrm{V}\otimes\mathbf{V}_\mathrm{H}$,
$$\mathbf{U}=\left[\mathbf{e}_{M_\mathrm{u}}\left(\gamma_{1}\right), \mathbf{e}_{M_\mathrm{u}}\left(\gamma_{2}\right), \cdots, \mathbf{e}_{M_\mathrm{u}}\left(\gamma_{M_\mathrm{u}}\right) \right] \in \mathbb{C}^{{M_\mathrm{u}}\times {M_\mathrm{u}}},$$ $$\mathbf{V}_\mathrm{H}=\left[\mathbf{e}_{N_H}\left(\varphi_{1}\right), \mathbf{e}_{N_H}\left(\varphi_{2}\right), \cdots, \mathbf{e}_{N_H}\left(\varphi_{N_\mathrm{H}}\right) \right] \in \mathbb{C}^{N_\mathrm{H}\times N_\mathrm{H}}$$ and $$\mathbf{V}_\mathrm{V}=\left[\mathbf{e}_{N_V}\left(\vartheta_{1}\right), \mathbf{e}_{N_V}\left(\vartheta_{2}\right), \cdots, \mathbf{e}_{N_V}\left(\vartheta_{N_\mathrm{V}}\right) \right] \in \mathbb{C}^{N_\mathrm{V}\times N_\mathrm{V}}$$ are unitary matrices. $\tilde{\mathbf{H}}_{i}^{c}$ denotes the \textit{beam domain channel matrix} with $\mathbf{e}_{M_\mathrm{u}}\left(\gamma_{m}\right)$ and $\mathbf{e}_{\mathrm{RIS}}\left(\varphi_{i_\mathrm{H}\left(n\right)},\vartheta_{i_\mathrm{V}\left(n\right)}\right)$ representing the transmit and receive beam, which can be calculated using the following approximation 
\begin{equation}\label{approximation}
\left[\tilde{\mathbf{H}}_{i}^{c}\right]_{nm}\approx \sum\limits_{p\in\mathbb{S}_{r,m}\cap\mathbb{S}_{t,n}}a_{i,p}^{c}e^{-j2\pi d_{i,p}^{c}/\lambda}
\end{equation}
where $\mathbb{S}_{t,m}$ is the set of all paths whose angles of departure are nearest to the sampling angle $\gamma_{m}$, and $\mathbb{S}_{r,n}$ is the set of paths whose angles of arrival are nearest to the sampling angles $\vartheta_{i_{\mathrm{H}}\left(n\right)}$ and $\varphi_{i_\mathrm{V}\left(n\right)}$.\\
\indent From \eqref{approximation}, it can be seen that in the beam domain, different elements of $\tilde{\mathbf{H}}_{i}^{c}$ represent the signals transmitted over different transmit and receive sampling angles. Different from the summation of all path signals in the physical domain, the beam domain channel can separate the paths of different angles by different beams. \\
\indent When the user lies in rich scattering environment, each element of $\tilde{\mathbf{H}}_{i}^{c}$ is the superposition of the attenuations of many paths. Thus the elements of $\tilde{\mathbf{H}}_{i}^{c}$ subject to the Gaussian distribution according to the central limit theorem, and the correlated beam domain channel model is in line with the Weichselberger model in~\cite{weichselberger2006stochastic}.\\
\indent The channel covariance matrices at the RIS and the $i$th user in cluster $c$ are given by
\begin{equation}\label{R_RIS}
\mathbf{R}_{\mathrm{RIS},i}^{c}=\mathbb{E}\left\{\mathbf{H}_{i}^{c}\left(\mathbf{H}_{i}^{c}\right)^\mathrm{{H}}\right\}=\mathbf{V}\mathbf{\Lambda}_{\mathrm{RIS},i}^{c}\mathbf{V}^\mathrm{H},
\end{equation}
and
\begin{equation}
\mathbf{R}_{\mathrm{user},i}^c=\mathbb{E}\left\{\left(\mathbf{H}_{i}^{c}\right)^\mathrm{{H}}\mathbf{H}_{i}^{c}\right\}=\mathbf{U}\mathbf{\Lambda}_{\mathrm{user},i}^{c}\mathbf{U}^\mathrm{H},
\end{equation}
where $\mathbf{\Lambda}_{\mathrm{RIS},i}^{c}=\text{diag}\left(\sum_{p\in\mathcal{S}_{t,1}}\beta_{i,p}^{c}, \cdots, \sum_{p\in\mathcal{S}_{t,N}}\beta_{i,p}^{c}\right)$ and $\mathbf{\Lambda}_{\mathrm{user},i}^{c}=\text{diag}\left(\sum_{p\in\mathcal{S}_{r,1}}\beta_{i,p}^{c}, \cdots, \sum_{p\in\mathcal{S}_{r,M_{\mathrm{u}}}}\beta_{i,p}^{c}\right)$ describe the power distribution over beams and {depend on the channel power angle spectrum (PAS)} $f_{\mathrm{RIS},i}^{c}\left(\varphi,\vartheta\right)$ at the RIS and $f_{\mathrm{user},i}^{c}\left(\gamma\right)$ at the $i$th user in cluster $c$, respectively~\cite{you2015pilot}.\\
\indent Since users in the same cluster share a similar scattering environment, it is reasonable to assume users in the same cluster have the same channel PAS. Then, the channel covariance matrices at the RIS and the users in cluster $c$ are given by
\begin{equation}
\mathbf{R}_{\mathrm{RIS}}^{c}=\mathbf{V}\mathbf{\Lambda}_{\mathrm{RIS}}^{c}\mathbf{V}^\mathrm{H},
\end{equation}
and
\begin{equation}
\mathbf{R}_{\mathrm{user}}^c=\mathbf{U}\mathbf{\Lambda}_{\mathrm{user}}^{c}\mathbf{U}^\mathrm{H},
\end{equation}
where $\mathbf{\Lambda}_{\mathrm{RIS}}^{c}$ and $\mathbf{\Lambda}_{\mathrm{user}}^{c}$ describe the power distribution over beams at the RIS and users in cluster $c$, respectively.

\subsection{Channel Model for the RIS-BS Channel}
In practice, the BS and RIS are usually deployed above surrounding objects. This means RIS-BS channels turn to be sparse for limited scattering objects and can be modeled as geometry channels with several dominant propagation paths. The channel between them can be modeled as
\begin{equation}\label{G}
\begin{aligned}
\mathbf{G}=\sqrt{\frac{M N}{L}} \sum_{l=1}^{L} \rho_{l} \mathbf{e}_M\left(\gamma_{l}\right) \mathbf{e}_\text{RIS}^{\mathrm{H}}\left(\varphi_{l}, \vartheta_{l}\right),
\end{aligned}
\end{equation}
where $\gamma_{l}$, $\varphi_{l}$ and $\vartheta_{l}$ represent AoA associated with the BS, azimuth AoD and elevation AoD associated with the RIS,  respectively. $L$ is the number of dominant paths, $\rho_{l}$ denotes the gain of the $l$th path.\\
\indent The BS-RIS channel is quasi-static since both the locations of BS and RIS are fixed~\cite{wu2019towards}. It only need to be estimated in a large timescale. The quasi-static RIS-BS channels can be estimated using methods in~\cite{hu2021two,he2019cascaded}. Therefore, in this paper, we assume $\mathbf{G}$ is known at the BS and focus on the estimation of channels between RIS and users, which is the main challenge in channel estimation for RIS-assisted systems
\subsection{For Single Antennal Users}
This paper considers the system setup where users are equipped with a single antenna and the system multiplexing gain is achieved due to multiple users. Under this circumstance, the received signal of the BS at time $t$ is expressed as
\begin{equation}
\mathbf{y}_t=\sum_{c=1}^{C}\sum_{i=1}^{{I}_c} \mathbf{G} \mathbf{\Phi}_{t} \mathbf{h}_{i}^{c} {s}_{ti}^{c} +\mathbf{n}_{t},
\end{equation}
where $\mathbf{h}_{i}^{c}\in \mathbb{C}^{N \times 1}$ represents the channel from the $i$th user in cluster $c$ to the RIS. ${s}_{ti}^{c}$ represents the transmitted symbol from the $i$th user in cluster $c$ at time $t$.\\
\indent Based on (16), the spatially correlated Rayleigh fading channel between the RIS and user $i$ in cluster $c$ is
\begin{equation}
\mathbf{h}_{i}^{c}\sim\mathcal{C N}\left(\mathbf{0}, \mathbf{R}_{i}^{c}\right),c\in\left\{1,2,\dots,C\right\},i\in\left\{1,2,\dots,I_{c}\right\},
\end{equation}
where $\mathbf{R}_{i}^{c}$ is non-diagonal and denotes the covariance matrix of the channel between RIS and user $i$ in cluster $c$. The covariance matrix $\mathbf{R}_{i}^{c}$ depends on the wavelength of transmitted signals and the scattering environment corresponding to cluster $c$~\cite{bjornson2020rayleigh}. Therefore, similar to (18), channels between RIS and users in the same cluster have the same covariance matrix, i.e.,
\begin{equation}
\mathbf{R}_{1}^{c}=\mathbf{R}_{2}^{c}=\cdots=\mathbf{R}_{I_c}^{c}=\mathbf{R}_{c}, c\in\left\{1,2,\dots,C\right\},
\end{equation}
where $\mathbf{R}_{c}$ is the covariance matrix of channesl between RIS and users in cluster $c$.\\
\indent From~\eqref{R_RIS}, it can been seen that the eigenvectors of the channel covariance matrix are determined by array response vectors of the RIS. Thus, the channel covariance matrices of users served by the same RIS share the same set of eigenvectors. The covariance matrix of $\mathbf{h}_{i}^{c}$ can be expressed in the form of eigenvalue decomposition
\begin{equation}\label{R_c}
\mathbf{R}_{c}=\mathbf{V}\mathbf{\Lambda}_{c}\mathbf{V}^\mathrm{H},c\in\left\{1,2,\dots,C\right\},
\end{equation}
where $\mathbf{V}\in \mathbb{C}^{N \times N}$ is the unitary matrix which is identical for all users. Ignoring the small eigenvalues~\cite{bjornson2009framework}, the diagonal matrix $\mathbf{\Lambda}_{c}$ consists of $E_{c}$ eigenvalues. The eigenvectors associated with the $E_{c}$ eigenvalues span the eigenspace where channel realizations of $\mathbf{h}_{i}^{c}, i\in\left\{1,2,\dots,I_{c}\right\}$ reside, which can be expressed as
\begin{equation}\label{S_c}
\mathcal{S}_c=\text{span}\left(\mathbf{d}_{f\left(c,1\right)},\mathbf{d}_{f\left(c,2\right)},\cdots,\mathbf{d}_{f\left(c,E_c\right)}\right),c\in\left\{1,2,\dots,C\right\},
\end{equation}
where $f\left(c,j\right)=\sum_{i=0}^{c-1}E_i+j$ and $E_0=0$, $\mathbf{d}_{f\left(c,j\right)}$ is the eigenvector associated with $\lambda_{f\left(c,j\right)}$. $\lambda_{f\left(c,j\right)}$ is the $j$th non-zero eigenvalue of $\mathbf{R}_{c}$, $j \in\left\{1,2,\dots,E_c\right\}$.

\begin{figure}[!htbp]
\centering
\includegraphics[width=3in]{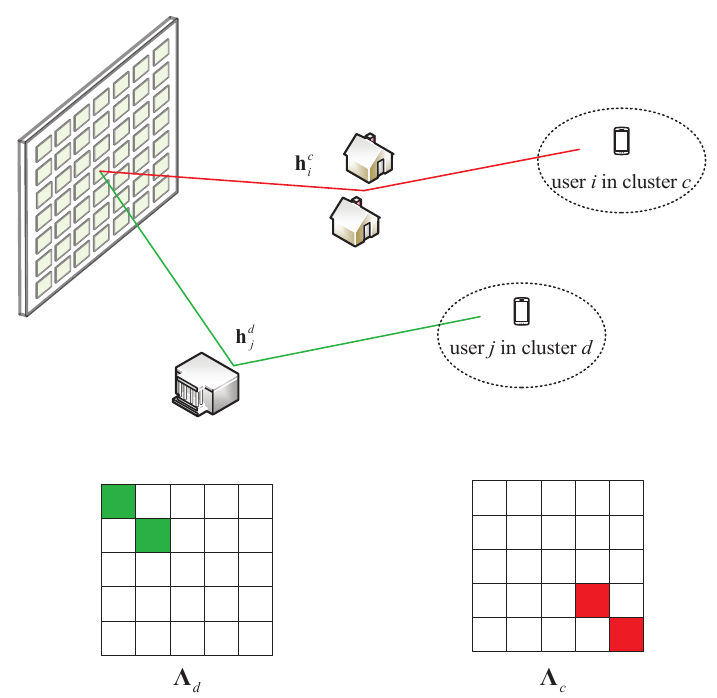}
\caption{An example of RIS-user channels in different clusters reside in different eigenspace.}
\label{eigenspace}
\end{figure}

In realistic outdoor wireless propagation environments where the RIS is located at an elevated position, most of the channel power is concentrated in a narrow angle interval, which indicates most of the channel power lies in a limited number of spatial directions (beams)~\cite{clerckx2013mimo}. For users located geographically apart in different clusters, the overlaps of their channel power in the angular domain might be neglected. As shown in Fig.~\ref{eigenspace}, it is assumed that the channel power seen at the RIS from the users in cluster $c$ and $d$, $\forall c,d \in\left\{1,2,\cdots,C\right\}, c\ne d $, is constrained in the AoA interval $\mathcal{A}_{i}=\left[\varphi_{\mathrm{min}}^{i},\varphi_{\mathrm{max}}^{i} \right]\times\left[\vartheta_{\mathrm{min}}^{i},\vartheta_{\mathrm{max}}^{i}\right], i\in\left\{c,d\right\}$ and $\mathcal{A}_{c}$ and $\mathcal{A}_{d}$ are non-overlapping. Recall that $[\mathbf{\Lambda}_i]_{nn}$ describes the power distribution over the RIS receiving beam $\mathbf{d}_n, n\in\left\{1,2,\cdots,N\right\}$, $i\in\left\{c,d\right\}$. Provided that AoA intervals , we have $\mathbf{\Lambda}_c\odot\mathbf{\Lambda}_d=\mathbf{0}$. This means that the RIS-user channels for users in cluster $c$ and $d$ reside in orthogonal eigenspace. In this circumstance, the BS can employ statistical CSI to distinguish users from different clusters and thus pilot reuse among clusters becomes feasible and beneficial.
\section{Uplink Pilot Transmission and Pilot Reuse Scheme}
\begin{figure}[!htbp]
\centering
\includegraphics[width=3in]{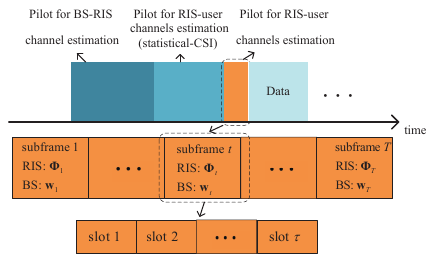}
\caption{The proposed channel estimation framework.}
\label{framework}
\end{figure}
\begin{figure}[!htbp]
\centering
\includegraphics[width=3in]{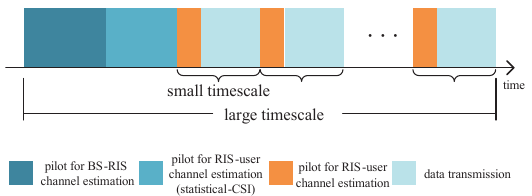}
\caption{Details of the proposed pilot transmission.}
\label{Details}
\end{figure}

The proposed channel estimation framework is exhibited in Fig.~\ref{framework}. Since the BS-RIS channel and the statistical CSI of RIS-user channels vary less frequently than the instantaneous CSI of RIS-user channels, they are estimated in a large timescale $T_L$, while the instantaneous CSI of RIS-user channels are estimated based on the uplink pilots in a small timescale $T_S$, and $T_L=\alpha T_S$ with $\alpha \gg 1$.\\
\indent For large timescale channel estimation, the quasi-static RIS-BS channel is assumed to be known by the BS using method in~\cite{hu2021two}. The BS estimates the statistical CSI of RIS-user channels by forming sample correlation matrices, which will be explained later in section IV.\\
\indent For small timescale channel estimation, as shown in Fig.~\ref{Details}, the uplink pilot transmission frame consists of $T$ subframes, and each subframe lasts for $\tau =\text{max}\left(I_1,I_2,\cdots,I_C\right)$ time slots. To better illustrate our idea, it is assumed that all the cluster contains the same number of users, i.e., $I_1=I_2=\cdots=I_C=I$. During the $\tau$ time instants in a subframe, the users transmit orthogonal uplink pilot sequences, i.e.,
\begin{equation}
\begin{aligned}
&\mathbf{x}_{\pi}  \in \mathbb{C}^{\tau \times 1}, \pi \in\left\{ 1, 2, \cdots ,\tau\right\},\\
&\mathbf{x}_{\pi_i}^\mathrm{H}\mathbf{x}_{\pi_j}=\tau P\delta\left(\pi_i-\pi_j\right),  \pi_i, \pi_j \in\left\{1,2,\dots,\tau\right\}
\end{aligned}
\end{equation}
with $P$ denoting the transmit power of each user. Specifically, pilot sequence $\mathbf{x}_{\pi}$ is assigned to the $\pi$th user in each cluster, $\pi \in\left\{ 1, 2, \cdots ,\tau\right\}$. This means that all the users in the same cluster utilize different pilot sequences, while the $\pi$th user in each cluster utilize the same pilot sequence, $\pi \in\left\{ 1, 2, \cdots ,\tau\right\}$. {The proposed pilot reuse strategy can reduce the pilot overhead for estimating the RIS-user channels in the small timescale by a factor of $C$ since the pilot sequences are reused among clusters.}\\
\indent The received pilots at the BS at subframe $t$, $t \in\left\{1,2,\dots,T\right\}$, is
\begin{equation}
\mathbf{Y}_t=\sum_{c=1}^{C}\sum_{i=1}^{\tau} \mathbf{G} \mathbf{\Phi}_{t} \mathbf{h}_{i}^{c} \mathbf{x}_{i}^\mathrm{T} +\mathbf{N}_{t},
\end{equation}
where $\mathbf{\Phi}_{t}$ is the reflection coefficient matrix at the RIS in subframe $t$, $t\in\left\{ 1, 2, \cdots ,T\right\}$. $\mathbf{N}_{t}\in \mathbb{C}^{M \times \tau}$ represent the noise at the BS. Each column of $\mathbf{Y}_t\in \mathbb{C}^{M \times \tau}$ is the received pilots in a single time slot. \\
\indent Then, for users assigned with pilot sequence $\mathbf{x}_{\pi}$, by right multiplying the conjugate of the pilot sequence, the BS obtains the channel observation as
\begin{equation}
\begin{aligned}
\tilde{\mathbf{y}}_{t,\pi}&=\frac{1}{\tau P}\mathbf{Y}_t\mathbf{x}_{\pi}^{*}\\
&=\sum_{c=1}^{C} \mathbf{G} \mathbf{\Phi}_{t} \mathbf{h}_{\pi}^{c}  +\frac{1}{\tau P}\mathbf{N}_{t}\mathbf{x}_{\pi}^{*}\\
&= \mathbf{G} \mathbf{\Phi}_{t} \mathbf{h}_{\pi}^{\mathrm{sum}}  +\tilde{\mathbf{n}}_{t,\pi},
\end{aligned}
\end{equation}
where $\mathbf{h}_{\pi}^{\mathrm{sum}}=\sum_{c=1}^{C}\mathbf{h}_{\pi}^{c}$ is the sumation of channels of users assigned with pilot sequence $\mathbf{x}_{\pi}$. $\tilde{\mathbf{n}}_{t,\pi}=\frac{1}{\tau P}\mathbf{N}_{t}\mathbf{x}_{\pi}^{*}\sim\mathcal{C N}\left(0, \frac{\sigma^2}{\tau P}\mathbf{I}\right)$.\\
\indent For a given $\pi$, $\pi \in\left\{1,2,\dots,\tau\right\}$, the channel observations $\left\{\tilde{\mathbf{y}}_{t,\pi}|t=1, 2,\dots,T\right\}$ are dependent on channels between RIS and users assigned pilot sequence $\mathbf{x}_{\pi}$, but independent of channels between RIS and users assigned other pilot sequences. That is to say, the problem of estimating the channels between RIS and users can thus be decomposed into $\tau$ independent subproblems, where the $\pi$th subproblem is to estimate the channels between RIS and users assigned pilot sequence $\mathbf{x}_{\pi}$, $\pi \in\left\{1,2,\dots,\tau\right\}$.
\section{Proposed EP Channel Estimation Scheme}
In this section, the EP channel estimation algorithm is proposed to solve subproblems derived in the previous section.\\
\indent In the $\pi$th subproblem, $\pi \in\left\{1,2,\dots,\tau\right\}$, we aim to estimate RIS-user channels $\left\{\mathbf{h}_\pi^c|c=1, 2,\dots,C\right\}$ based on channel observations $\left\{\tilde{\mathbf{y}}_{t,\pi}|t=1, 2,\dots,T\right\}$. Assuming the RIS aided system employ linear combining at the BS, the combined signal at the BS at subframe $t$, $t \in\left\{1,2,\dots,T\right\}$, is
\begin{equation}
\begin{aligned}
r_{t,\pi}&=\mathbf{w}^\mathrm{H}\tilde{\mathbf{y}}_{t,\pi}\\
&=\mathbf{w}^\mathrm{H}\mathbf{G}\mathbf{\Phi}_t\mathbf{h}_{\pi}^\mathrm{sum}+z_{t,\pi},
\end{aligned}
\end{equation}
where $\mathbf{w}$ and $z_{t,\pi}=\mathbf{w}^\mathrm{H}\tilde{\mathbf{n}}_{t,\pi}\sim\mathcal{C N}\left(0, \frac{\sigma^2}{\tau P}\right)$ represent the normalized combing vector and the noise at the BS, respectively. \\
\indent The covariance matrix of the sum channels $\mathbf{h}_{\pi}^\mathrm{sum}$, can be expressed as
\begin{equation}
\mathbf{R}_\mathrm{sum}=\sum_{c=1}^C\mathbf{R}_{c}.
\end{equation}

The eigenspace of sum channel can be expressed as
\begin{equation}\label{S}
\begin{aligned}
\mathcal{S}&=\mathcal{S}_1\oplus\mathcal{S}_2\oplus\cdots\oplus\mathcal{S}_C\\
&=\text{span}\left(\mathbf{d}_{1},\mathbf{d}_{2},\cdots,\mathbf{d}_{E}\right),
\end{aligned}
\end{equation}
where $E=\sum_{k=c}^C{E_c}$ is the dimension of eigenspace $\mathcal{S}$, $\left\{\mathbf{d}_{1},\mathbf{d}_{2},\cdots,\mathbf{d}_{E}\right\}$ is a set of orthonormal bases for eigenspace $\mathcal{S}$, i.e.,
\begin{equation}
\mathbf{d}_i^\mathrm{H}\mathbf{d}_j=\delta\left(i-j\right),  i, j \in\left\{1,2,\dots,E\right\}.
\end{equation}

Let the number of subframes in an uplink pilot transmission frame equals the dimension of eigenspace $\mathcal{S}$, i.e., $T=E$. The $ \mathbf{h}_{\pi}^\mathrm{sum}$ can be expressed by
\begin{equation}
\mathbf{h}_{\pi}^\mathrm{sum}=\eta_{1,\pi}\mathbf{d}_{1}+\eta_{2,\pi}\mathbf{d}_{2}+\cdots+\eta_{T,\pi}\mathbf{d}_{T},
\end{equation}
where $\eta_{t,\pi}\sim\mathcal{C N}\left(0, \lambda_t\right), t \in\left\{1,2,\dots,T\right\}$, is the projection parameter which denotes the projection of $\mathbf{h}_{\pi}^\mathrm{sum}$ to the eigenvector $\mathbf{d}_{t}$, i.e., $\eta_{t,\pi}=\mathbf{d}_{t}^\mathrm{H}\mathbf{h}_{\pi}^\mathrm{sum}$.

Based on the RIS-BS channel $\mathbf{G}$ and the pilot signal at time instant $t$, the BS combining vector and the combined signal at time instant $t$, $t \in\left\{1,2,\dots,T\right\}$, are given by
\begin{equation}\label{w}
\mathbf{w}=\mathbf{u}
\end{equation}
and 
\begin{equation}\label{r}
\begin{aligned}
r_{t,\pi}&=\delta{\mathbf{v}}^\mathrm{H}\mathbf{\Phi}_t\mathbf{h}_{\pi}^\mathrm{sum}+z_{t,\pi}\\
&=\delta\tilde{\mathbf{d}}_t^\mathrm{H}\mathbf{h}_{\pi}^\mathrm{sum}+z_{t,\pi},
\end{aligned}
\end{equation}
where $\tilde{\mathbf{d}}_t=\mathbf{\Phi}_t^\mathrm{H}\mathbf{v}$, $\delta$ denotes the largest singular value derived from the singular value decomposition (SVD) of RIS-BS channel, i.e., $\mathbf{G}=\mathbf{U}_{G} \mathbf{\Sigma} \mathbf{V}_{G}^\mathrm{H}$. $\mathbf{u}$ and $\mathbf{v}$ are the columns of unitary matrices $\mathbf{U}_{G}$ and $\mathbf{V}_{G}$ corresponding to $\delta$, respectively. 

{Set the RIS reflection coefficient matrix $\mathbf{\Phi}_t$, $t \in\left\{1,2,\dots,E\right\}$, as 
\begin{equation}\label{theta}
\begin{aligned}
&{a}_{t,n}=\kappa_t\left\lvert \frac{{d}_{t,n}}{{v}_{n}} \right\rvert,n\in\left\{1,2,\dots,N\right\},\\
&{\theta}_{t,n}=-\text{arg}\left(\frac{{d}_{t,n}}{{v}_{n}}\right),n\in\left\{1,2,\dots,N\right\},
\end{aligned}
\end{equation}
where $\kappa_t$ is the scale factor to make sure the reflection amplitude of REs is not greater than $1$. ${d}_{t,n}$ and ${v}_{n}$ are the $n$th elements of $\mathbf{d}_t$ and $\mathbf{v}$, respectively. We have
\begin{equation}\label{d_tilde}
\tilde{\mathbf{d}}_t= \kappa_t\mathbf{d}_t.
\end{equation}}
\indent Substituting~\eqref{d_tilde} into~\eqref{r}, the combined signal at BS at time instant $t$, $t \in\left\{1,2,\dots,T\right\}$, is given by
\begin{equation}
r_{t,\pi}=\delta\kappa_t\eta_{t,\pi}+z_{t,\pi}.
\end{equation}
\indent {Using the minimum mean square error (MMSE) estimator, the estimation of $\eta_{t,\pi}$ , $t\in\left\{1,2,\dots,T\right\}$, is
\begin{equation}\label{MMSE}
\begin{aligned}
\hat{\eta}_{t,\pi}(r_{t,\pi}) &=\mathbb{E}\{\eta_{t,\pi}|r_{t,\pi}\}\\
&=\frac{P\tau\delta\kappa_t\lambda_t}{P\tau\delta^2\kappa_t^2\lambda_t+\sigma^2}r_{t,\pi}.
\end{aligned}
\end{equation}}
The estimation of the sum channel $\mathbf{h}_{\pi}^\mathrm{sum}$ is
\begin{equation}\label{h_hat}
\begin{aligned}
\hat{\mathbf{h}}_{\pi}^\mathrm{sum} = \sum_{t=1}^E\hat{\eta}_{t,\pi}\mathbf{d}_{t}.
\end{aligned}
\end{equation}

Finally, the estimation of the channel between the RIS and the $\pi$th user in cluster $c$, $ c\in\left\{1,2,\dots,C\right\}$, is given by dividing $\hat{\mathbf{h}}_{\pi}^\mathrm{sum}$ into the corresponding eigenspace,
\begin{equation}
\hat{\mathbf{h}}_{\pi}^{c}=\sum_{\mathbf{d}_{t}\subseteq\mathcal{S}_{c}, t\in\left\{1,2,\dots,T\right\}}\hat{\eta}_{t,\pi}\mathbf{d}_{t}.
\end{equation}

\floatname{algorithm}{Algorithm}
\renewcommand{\algorithmicrequire}{\textbf{Input:}}
\renewcommand{\algorithmicensure}{\textbf{Output:}}
\begin{algorithm}
\caption{The Proposed EP Channel Estimation Algorithm.}\label{algorithm1}
        \begin{algorithmic}[1] 
            \Require Received pilots $\left\{\mathbf{Y}_t|t=1, 2, \cdots, T\right\}$, channel covariance information $\left\{\mathbf{R}_c|c=1, 2, \cdots, C\right\}$, orthogonal pilots $\left\{\mathbf{x}_\pi|\pi=1, 2, \cdots, \tau\right\}$ and the BS-RIS channel $\mathbf{G}$.
            \Ensure The estimation channels bwtween the RIS and users $\{\hat{\mathbf{h}}_i^c|c=1, 2, \cdots, C, i=1, 2, \cdots, I_c\}$.
            \For{$ c=1:\tau $}
                \State Calculate $\mathcal{S}_{c}$ according to~\eqref{R_c} and~\eqref{S_c} 
             \EndFor
             \State Obtain $\mathcal{S}$ according to~\eqref{S} 
             \State Calculate $\mathbf{w}$ according to~\eqref{w} 
             \For{$ t=1:T $}
                 \State {Set $\mathbf{\Phi}_t$ according to~\eqref{theta}} 
                 \For{$ \pi=1:\tau $}
                     \State $ r_{t,\pi}=\frac{1}{\tau P}\mathbf{w}^\mathrm{H}\mathbf{Y}_t\mathbf{x}_{\pi}^{*} $
                     \State Estimate ${\eta}_{t,\pi}$ according to~\eqref{MMSE} 
                     \State Calculate $\hat{\mathbf{h}}_{\pi}^{c}$ according to~\eqref{h_hat} 
                   \EndFor
                \EndFor
             \State \Return{$\left\{\hat{\mathbf{h}}_i^c|c=1, 2, \cdots, C, i=1, 2, \cdots, I_c\right\}$}

        \end{algorithmic}
    \end{algorithm}

The proposed EP channel estimation algorithm is summarized in \textbf{Algorithm 1}. In Steps $1$-$4$, the eigenspace of sum channels is calculated based on the channel covariance information. In Step $5$, the combining vector at the BS is calculated based on RIS-BS channel $\mathbf{G}$. In Steps $6$-$13$, the RIS-user channel estimation problem is decomposed into $\tau$ independent subproblems, each of which estimates the channels between the RIS and users assigned pilot $\mathbf{x}_\pi$. Each subproblem is solved respectively using the EP method. Finally, the algorithm gives the estimation of channels between RIS and users.

\section{Pilot Overhead}
The quasi-static RIS-BS channel and the statistical CSI for RIS-user channels are estimated in a large timescale $T_L$. The quasi-static RIS-BS channel is estimated using the method in~\cite{hu2021two}, with the minimum required pilot overhead $\tau_1=2\left(N+1\right)$. The approach to estimate the correlation matrices for RIS-user channels is to form the sample correlation matrix. For users in cluster $c$, suppose the BS has made $\beta$ independent observations of $\mathbf{h}_{i}^{c}$. The correlation matrix for users in cluster $c$ is
\begin{equation}
{{\bf{R}}_{c}^{sample}} = \frac{1}{I_c\beta}\sum\limits_{i = 1}^{I_c}\sum\limits_{n = 1}^\beta {{\bf{h}}_{i,n}^{c}\left({\bf{h}}_{i,n}^c\right)^{\rm{H}}}, c\in \left\{1,2,\dots,C\right\}, 	
\end{equation}
where $\mathbf{h}_{i,n}^{c}$ denotes the $n$th observation of $\mathbf{h}_{i}^{c}$.\\
\indent For each element of ${\bf{R}}_{c}^{sample}$, the law of large numbers implies that the sample variance converges (almost surely) to the true variance of corresponding element of ${\bf{R}}_{c}$. The standard deviation of the sample variance decays as ${1 \mathord{\left/
 {\vphantom {1 {\sqrt {{I_c}\beta } }}} \right.
 \kern-\nulldelimiterspace} {\sqrt {{I_c}\beta } }}$, thus a small number of observations $\beta$ is sufficient to get a good variance estimate. {The observation of RIS-user channels can be obtained by using existing channel estimation method in~\cite{hu2021two}, which has the pilot overhead of $\left\lceil\frac{KN}{M}\right\rceil$ for our considered system. The required signaling overhead of obtaining all the correlation matrix ${\bf{R}}_{c}, c\in \left\{1,2,\dots,C\right\}$ is $\tau_2=\beta K\left\lceil {\frac{N}{M}} \right\rceil$.}\\
\indent From the time domain perspective, the channel covariance matrices vary more slowly than the instantaneous CSI, and thus can be estimated in a large timescale. From the frequency domain perspective, the channel covariance matrices have been shown to stay constant over a wide frequency interval~\cite{barriac2004space}, and thus can be estimated via averaging over frequency. Therefore, there will be enough time-frequency resources to estimate the channel covariance matrices, and the estimation accuracy can be guaranteed in practice.\\
\indent Then, in a small timescale $T_S$, the RIS-user channels are estimated based on the uplink pilots using the proposed EP algorithm. The required signaling overhead of obtaining all the RIS-user channels is $\tau_3=\frac{{KE}}{C}$.

During a time period of $T_S$, the average pilot overhead is calculated by
\begin{equation}
\begin{aligned}
\tau&=\frac{T_L}{T_S}\left(\tau_1+\tau_2\right)+\tau_3\\
&=\frac{{2\left( {N - 1} \right) + {\beta}K\left\lceil {\frac{N}{M}} \right\rceil }}{\alpha } + \frac{{KE}}{C}
\end{aligned}
\end{equation}

\begin{table}[!t]
\caption{Pilot Overhead Comparison of Different\\ Channel Estimation Schemes}\label{pilot_overhead}
\centering
\renewcommand{\arraystretch}{2}{
\begin{tabular}{|c||c|}
\hline
Channel Estimation Schemes & Required Pilot Overhead\\
\hline
LS Estimation~\cite{jensen2020optimal} & $NK$\\
\hline
Three-Phase Estimation~\cite{wang2020channel} & $N+\text{max}\left(K-1, \left\lceil {\frac{{\left( {K - 1} \right)N}}{M}} \right\rceil\right)$\\
\hline
Two-Timescale Estimation~\cite{hu2021two} & $\frac{{2\left( {N - 1} \right)}}{\alpha } + K\left\lceil {\frac{N}{M}} \right\rceil $\\
\hline
Proposed Scheme & $\frac{{2\left( {N - 1} \right) + {\beta}K\left\lceil {\frac{N}{M}} \right\rceil }}{\alpha } + \frac{{KE}}{C} $\\
\hline
\end{tabular}}
\end{table}
In Table~\ref{pilot_overhead}, the pilot overhead of the proposed channel estimation scheme is compared with the LS channel estimation~\cite{jensen2020optimal}, the three-phase channel estimation~\cite{wang2020channel} and the two-timescale channel estimation~\cite{hu2021two}.\\
\indent From Table~\ref{pilot_overhead}, it can be seen that the dimension of eigenspace, $E$, is a key factor that contributes to the pilot reduction in our proposed channel estimation scheme. By leveraging the correlated property of RIS-user channels, the pilot overhead for estimating RIS-user channels is proportion to $E$ in our method, while the corresponding pilot overheads for methods in~\cite{wang2020channel,hu2021two} are proportion to the number of RIS REs $N$.  When the BS-RIS channel varies much more slowly than the RIS-UE channel and BS-UE channel do, the advantage of the proposed method is significant. On the contrary, when the BS-RIS channel also varies fast, the proposed method will not achieve significant pilot reduction.
Besides, it can be seen that $\alpha$ is also important for the pilot reduction. When the BS-RIS channel and statistical CSI of RIS-user channels varies much more slowly than the instantaneous CSI of RIS-user channels $\left(\alpha\gg 1\right)$, the advantage of the proposed method compared with methods in~\cite{jensen2020optimal} and~\cite{wang2020channel} is significant.

\begin{figure}[!htbp]
\centering
\includegraphics[width=3.5in]{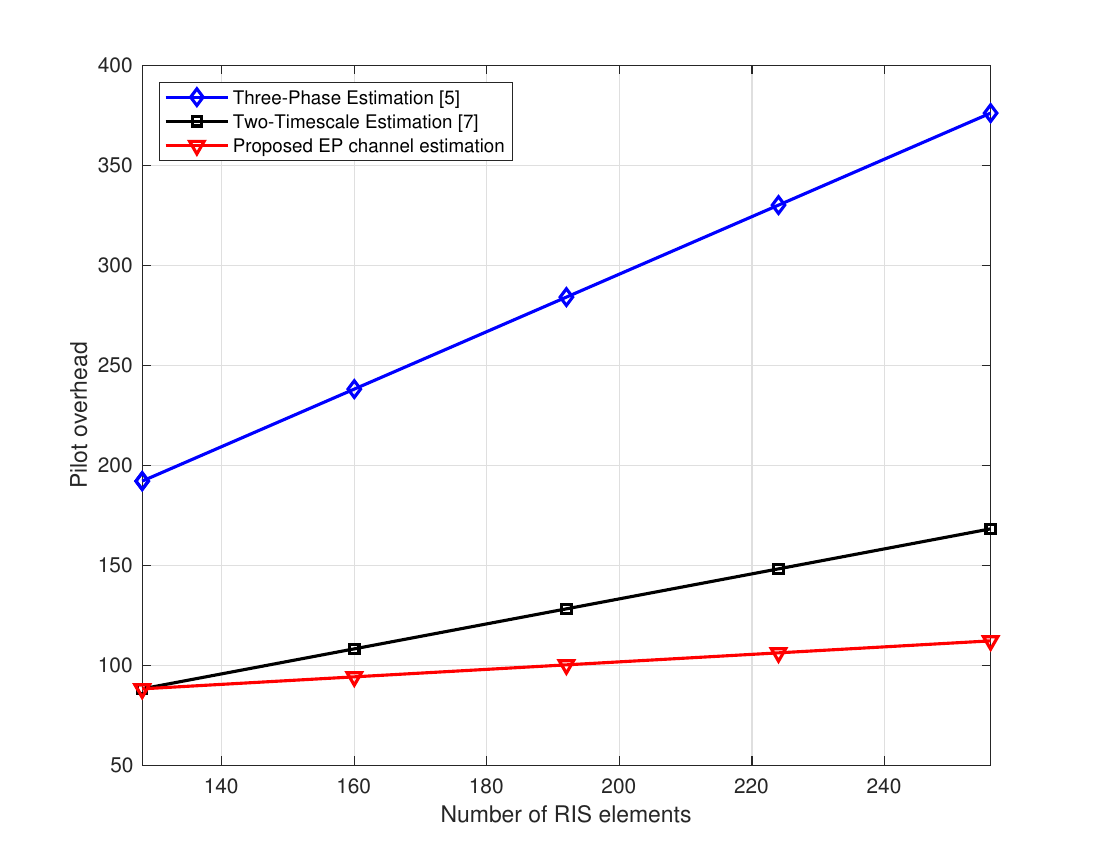}
\caption{The pilot overhead against $N$, where $M = 16$, $K = 8$.}\label{pilot_over_N}
\end{figure} 

\begin{figure}[!htbp]
\centering
\includegraphics[width=3.5in]{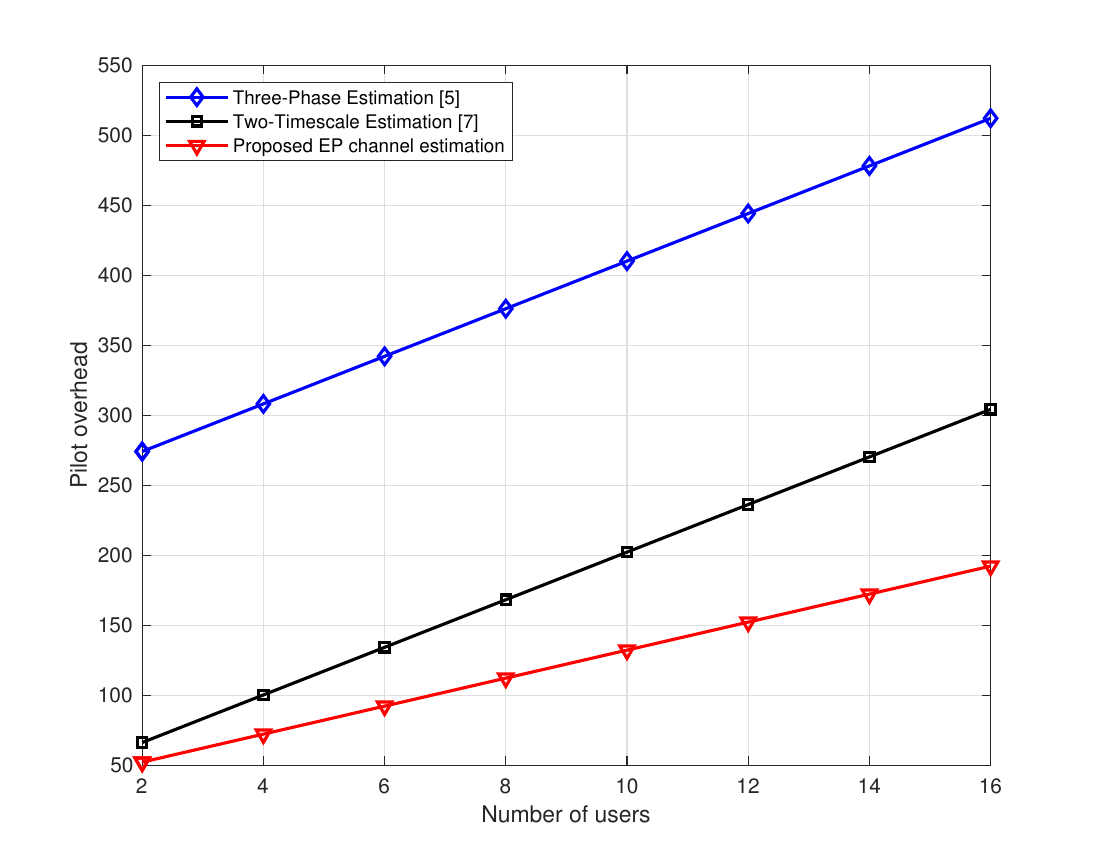}
\caption{The pilot overhead against $K$, where $M = 16$, $N = 256$.}\label{pilot_over_K}
\end{figure} 

\begin{figure}[!htbp]
\centering
\includegraphics[width=3.5in]{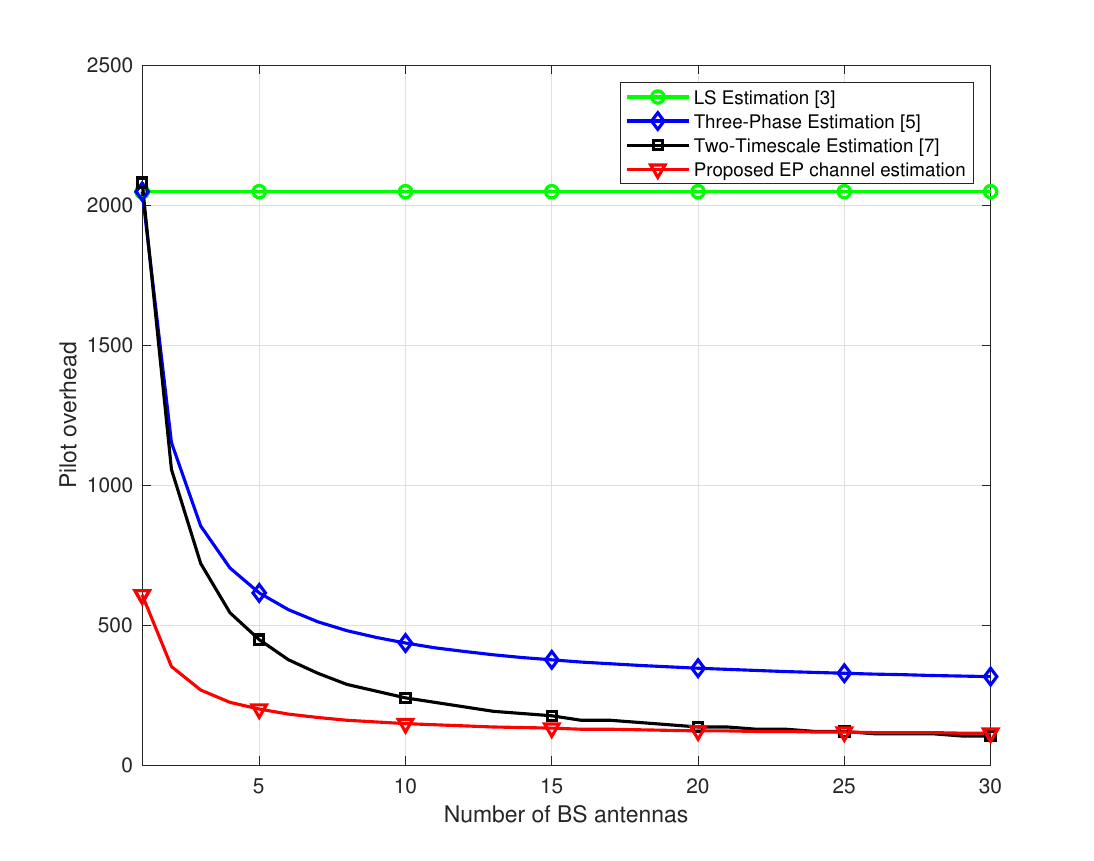}
\caption{The pilot overhead against $M$, where $N = 256$, $K = 8$.}\label{pilot_over_M}
\end{figure} 

{Fig.~\ref{pilot_over_N}-\ref{pilot_over_M} illustrate the relation between the pilot overhead and M, N, K, where $\alpha=16$, $\beta=4$, $C=4$, $E=32$.\\
\indent Fig.~\ref{pilot_over_N}-\ref{pilot_over_K} show that the pilot overheads for all schemes increase with the number of RIS REs and the users. In addition, it can be seen that the pilot overhead of the proposed EP channel estimation scheme is the lowest. It is because the proposed scheme leverages the correlated channel property and conducts channel estimation based on slow varying statistical CSI. Since the pilot overhead for the LS estimation is an order of magnitude higher than other schemes under all considered N(K), the LS estimation result is omitted.}\\
\indent In Fig.~\ref{pilot_over_M}, for the proposed channel estimation scheme and the schemes in~\cite{wang2020channel,hu2021two}, the pilot overhead decreases as the number of BS antennas increases because the pilot overhead for estimating the channel covariance information in the large timescale decreases.
\section{Simulations}
In this section, simulation results are provided to validate the effectiveness of the proposed algorithm and draw useful insights.
\subsection{ Simulation Setup}
\begin{figure}[!htbp]
\centering
\includegraphics[width=3in]{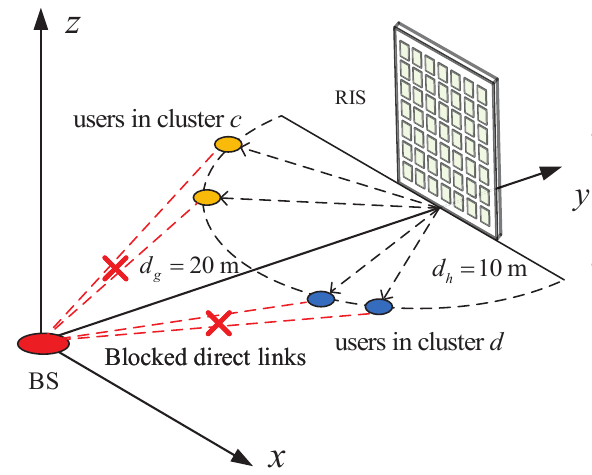}
\caption{The simulation setup.}\label{setup}
\end{figure} 
{As shown in Fig~\ref{setup}, we assume the BS is $20$ meters (m) away from the RIS. The users are located on a half-circle centered at the RIS with a radius of $10$ m.} In simulations, system parameters are set as follows unless otherwise specified: $M = 16$, $N=256$, $L=5$, $K = 8$,$C = 4$, $E=32$. The channel path-loss coefficients are modeled as $\rho_g=\rho_0\left(\frac{d_g}{d_0}\right)^{-\alpha_g}$ and $\rho_h=\rho_0\left(\frac{d_h}{d_0}\right)^{-\alpha_h}$, where $\rho_{0}=-20\mathrm{dB}$ is the path loss at reference distance $d_{0}=1\mathrm{m}$, $\alpha_h = 2.2$ and $\alpha_g = 2.1$.\\
\indent {For the RIS-BS channel, according to the channel model in~\eqref{G}, let $l=1$ represent the indices of the line-of-sight (LoS) components in $\mathbf{G}$. The complex channel gains are distributed as $\rho_1\sim \mathcal{C N}\left(0, 1\right)$ and $\rho_i\sim \mathcal{C N}\left(0, 10^{-0.5}\right), i\in\left\{2,\dots,L\right\}$. Angles $\varphi_l, \gamma_l$ and $\vartheta_l$ are uniformly generated from $[0,\pi]$.\\
\indent For the RIS-user channels, we consider the typical outdoor wireless propagation environments where the channel PAS can be modeled as the truncated Laplacian distribution~\cite{cho2010mimo}. The channel PAS for user $i$, $\forall i=1,2,\cdots,I_c$, in cluster $c$ is modeled as 
\begin{equation}
f_{\mathrm{RIS},i}^{c}\left(\varphi_c,\vartheta_c\right)=f_{\text{Lap}}\left(\varphi_c,\sigma^\varphi_c\right)f_{\text{Lap}}\left(\vartheta_c,\sigma^\vartheta_c\right), 
\end{equation}
where $\varphi_c\left(\vartheta_c\right)$ and $\sigma_c^{\varphi}\left(\sigma_c^{\vartheta}\right)$ represent the azimuth (elevation) angular standard deviation (ASD) and the azimuth (elevation) mean AoA of user channels in cluster $c$. $f_{\text{Lap}}\left(\varphi,\sigma^\varphi\right)$ represents the truncated Laplacian distribution with mean $\varphi$ and standard deviation (SD) $\sigma^\varphi$~\cite{molisch2012wireless}. We assume that $\sigma^\varphi_c=14^{\circ}$, $\sigma^\vartheta_c=2^{\circ}$, $\vartheta_c=20^{\circ}$, $\forall c=1,2,\cdots,4$,~\cite{bjornson2017massive} and the mean azimuth AoAs of the cluster $1$ to cluster $4$ are $\left[0.9273, 1.3694, 1.7722, 2.2143\right]$ in radians.}\\
\indent For the uplink pilot transmission, the signal-to-noise ratio is defined by
\begin{equation}
\mathrm{SNR}=\frac{P\rho_g\rho_h}{\sigma^2}.
\end{equation}
The normalized mean square error (NMSE)  
\begin{equation}
\frac{1}{K}\sum_{k=1}^{K}\mathbb{E}\left\{\frac{\left\lVert\mathbf{G}_k-\hat{\mathbf{G}}_k\right\lVert_\mathrm{F}^2}{\left\lVert\mathbf{G}_k\right\lVert_\mathrm{F}^2}\right\}
\end{equation}
is adopted as the performance metric, where $\mathbf{G}_k=\mathbf{G}\text{diag}\left({\mathbf{h}_k}\right)$ is the cascaded channel. All results are averaged over $10^{5}$ independent channel realizations.\\
\indent To verify the effectiveness of the proposed EP algorithm, its performance is compared with the channel estimation schemes from Table~\ref{pilot_overhead}.
\subsection{ Simulation Result}
\begin{figure}[!htbp]
\centering
\includegraphics[width=3.5in]{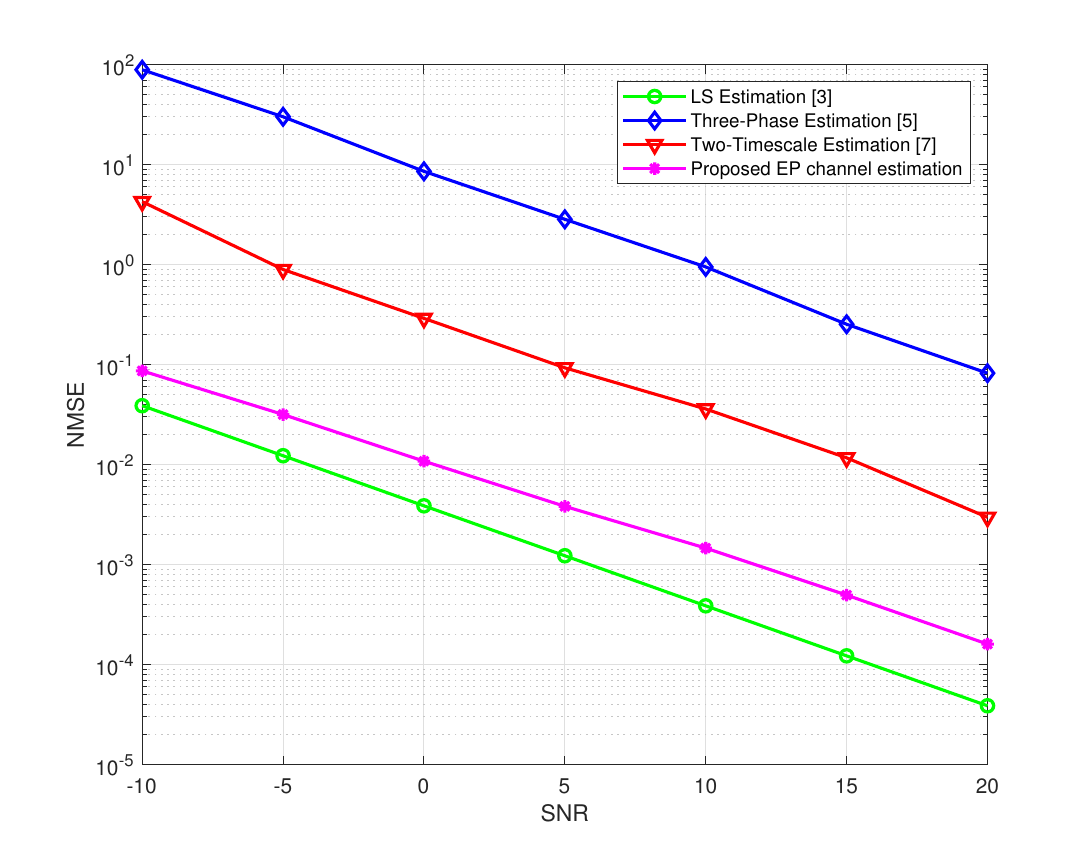}
\caption{SNR versus NMSE.}\label{NMSE_over_SNR}
\end{figure}
Fig.~\ref{NMSE_over_SNR} shows the NMSE of the cascaded channel against the SNR. The proposed channel estimation method can achieve lower NMSE than that in~\cite{wang2020channel,hu2021two}. Though the LS channel estimation in~\cite{jensen2020optimal} is more accurate, proposed method requires much fewer training pilots because it exploits statistical CSI of correlated channels between RIS and users as a prior.
\begin{figure}[!htbp]
\centering
\includegraphics[width=3.5in]{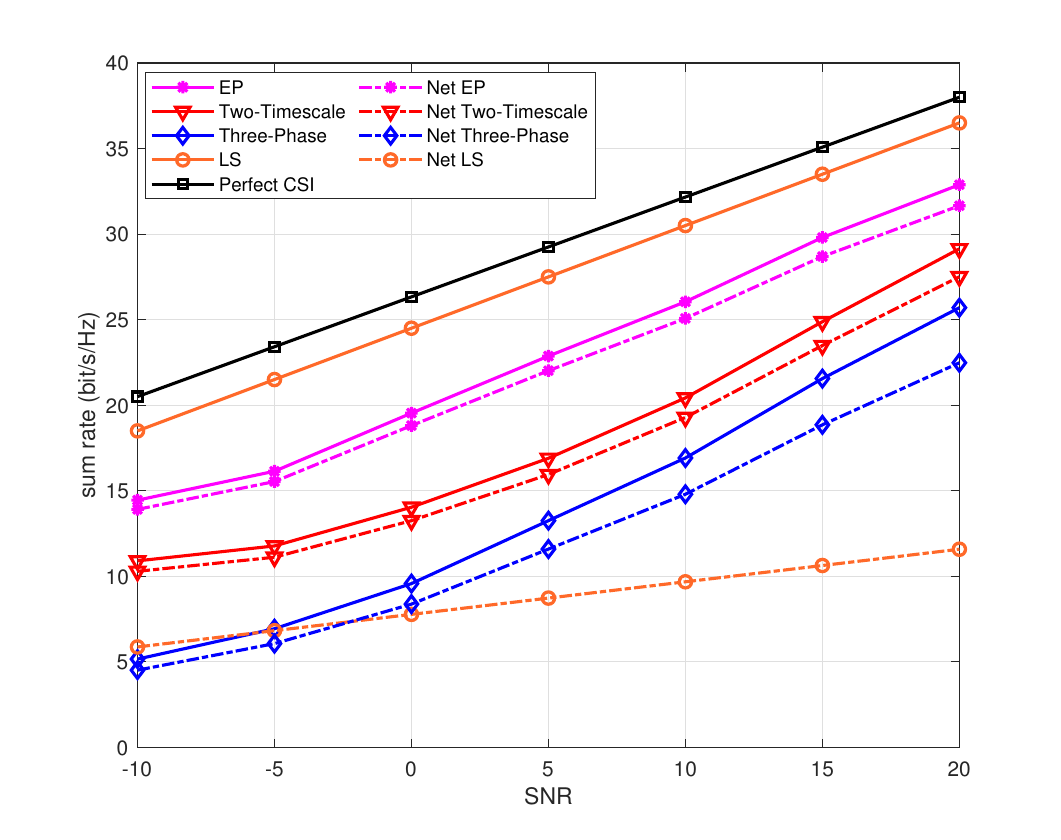}
\caption{SNR versus sum rate.}\label{rate_over_SNR}
\end{figure}

Fig.~\ref{rate_over_SNR} shows the downlink sum rate of the RIS-aided communication system using different channel estimation schemes. The joint precoding scheme in~\cite{wu2019intelligent} is adopted to jointly optimize the precoding matrix and the reflection coefficient matrix. The sum rate is calculated by
\begin{equation}
R_{\text{sum}}=\sum_{k=1}^{K}\text{log}\left(1+\text{SINR}_k^\mathrm{d}\right),
\end{equation}
where $\text{SINR}_k^\mathrm{d}$ denotes the signal-to-interference-plus-noise ratio for the $k$th user in the downlink. {From Fig.~\ref{rate_over_SNR}, we can observe that the channel estimation method with smaller NMSE has better performance for data transmission. The reason behind this is that inaccurate CSI results in inefficient joint beamforming design and thus degrades the system performance for data transmission.} In line with Figs.~\ref{NMSE_over_SNR}, it can be observed that the RIS-aided communication system adopts the proposed EP channel estimation scheme can achieve higher sum rate than the RIS-aided communication system that adopts scheme in~\cite{wang2020channel} and~\cite{hu2021two}. Although the performance is inferior to LS method in~\cite{jensen2020optimal}, the proposed scheme requires significantly lower pilot overhead. {To further clarify the domination of the proposed EP channel estimation scheme. The net sum rates of the RIS-aided communication system using different channel estimation schemes are drawn in dash lines in Fig.~\ref{rate_over_SNR}. The net sum rate is calculated by $R_{\text{sum}}^{\text{net}}=\frac{\tau_c-\tau_p}{\tau_c}R_{\text{sum}}$, with $\tau_c$ and $\tau_p$ representing the channel coherent time and the pilot length required by corresponding channel estimation scheme. We set $\tau_c=3000$ and $\tau_p$ can be cauculated according to Table~\ref{pilot_overhead}. It can be observed that the proposed EP scheme exhibits superior performance compared to the LS scheme in terms of net sum rate. This can be attributed to the lower pilot overhead required by the EP scheme.}

\section{Conclusion}
In this paper, a RIS-aided multi-user MISO communication system with clustered users is investigated. First, to describe the correlated feature for RIS-user channels, the beam domain channel model was developed for RIS-user channels. Then, a pilot reuse strategy was put forward to reduce the pilot overhead and decompose the channel estimation problem into several subproblems. Finally, by leveraging the correlated nature of RIS-user channels, an eigenspace projection (EP) algorithm was proposed with reduced pilot overhead to solve each subproblem respectively. Simulation results showed that the proposed EP channel estimation scheme can achieve accurate channel estimation with lower pilot overhead than existing schemes. 

\bibliographystyle{IEEEtran}
\bibliography{myref}

%
%
%
%
%

\vfill

\end{document}